\documentclass{JAC2003}

\usepackage{graphicx}
\usepackage{booktabs}

\begin{document}
\title{\uppercase{
Pulse-to-pulse Beam Modulation and \\
Event-based Beam Feedback System at KEKB Linac
}}

\author{K. Furukawa\thanks{\texttt{<\,kazuro.furukawa\,@\,kek.jp\,>}}, 
 M. Satoh, T. Suwada, and T.T. Nakamura, \\
 KEK, Tsukuba, Ibaraki, 305-0801, Japan
}

\maketitle

\begin{abstract}
 Beam injections to KEKB and Photon Factory are performed with pulse-to-pulse modulation at 50\,Hz. Three very different beams are switched every 20\,ms in order to inject those beams into KEKB HER, LER and Photon Factory (PF) simultaneously. Human operators work on one of those three virtual accelerators, which correspond to three-fold accelerator parameters. Beam charges for PF injection and the primary electron for positron generation are 50-times different, and beam energies for PF and HER injection are 3-times different. Thus, the beam stabilities are sensitive to operational parameters, and if any instability in accelerator equipment occurred, beam parameter adjustments for those virtual accelerators have to be performed. In order to cure such a situation, beam energy feedback system was installed that can respond to each of virtual accelerators independently.
\end{abstract}

\section{Introduction}

KEKB asymmetric electron positron collider has been operated for 10 years improving the experimental performance continuously.  At the beginning of the commissioning it was anticipated whether the positron injection efficiency was sufficient.  In order to enhance the injection efficiency several projects were performed, including double beam bunches in a pulse and continuous (top-up) injection during the experiment.  Such efforts improved the integrated luminosity, and the linac turned out not to be a bottleneck~\cite{lin-status-linac02}.  

Further challenge was that the injector is shared between four storage rings, namely, KEKB-HER (High Energy Ring), KEKB-LER (Low Energy Ring), PF (Photon Factory) and PF-AR (Photon Factory Advanced Ring), with quite different energies from 2.5\,GeV to 8\,GeV and charges from 0.1\,nC to 10\,nC as in Table~\ref{four-beams}.  There were concerns about the beam characteristic reproducibility on the beam mode changes~\cite{beamswitch-linac00}.

\begin{table}[hbt]
 \centering
 \caption{Typical beam modes}
 \begin{tabular}{lcccc}
  \toprule
  Injection & Beam & Energy & Charge & No. of \\ 
  Mode      &      &        & /Bunch & Bunches \\
  \midrule
  KEKB-HER & e$^-$ & 8.0\,GeV   & 1.2\,nC & 1 or 2 \\
  KEKB-LER & e$^+$ & 3.5\,GeV & 0.6\,nC & 2 \\
  KEKB-LER\,* & e$^-$ & 4.0\,GeV & 10\,nC & 2 \\
  PF       & e$^-$ & 2.5\,GeV & 0.1\,nC & 1 \\
  PF-AR    & e$^-$ & 3.0\,GeV & 0.1\,nC & 1 \\
  \bottomrule
 \multicolumn{5}{l}{* primary electron for positron generation.}
 \end{tabular}
 \label{four-beams}
\end{table}

For this reason, during the commissioning, many slow closed feedback loops were installed in order to stabilize the beam energies and orbits at several different locations along the 600-m linac~\cite{lin-efb-ical99}.  

Recently, we succeeded in pulse-to-pulse beam mode changes, with modulating many parameters in 20\,ms.  This is called as a simultaneous injection and it enabled both the luminosity tuning stability at KEKB and the top-up injection to PF~\cite{iuc-linac08,eventcont-ical09,simul-pac09}.   

It is considered whether it may improve the injection stability with beam feedback loops even under simultaneous injection.


\section{Slow Beam Feedback Loops}

Because beam instabilities were observed at the beginning of the linac commissioning for KEKB, many closed loops were installed.  If the energy fluctuated, the beam position, where the dispersion function was large, was measured and RF phases at two adjacent RF sources were modulated to opposite directions from the crest phase in order to maintain the energy but not to make energy spread.  If the orbit was unstable, beam positions at two locations where betatron phases are 90-degree apart were measured and corresponding steering coils were adjusted.  

In order to suppress the measurement noises, a weighed average over several locations were often used, instead of a single beam position information, based on the response functions to the energy change or the beam kick.  If the energy measurement noise was large, the betatron oscillation was measured using the beam positions at two locations where the dispersion functions were zero, and the information was used to compensate the energy displacement information. 

Those procedures were described employing a scripting computer language and operators could manipulate their parameters via a graphical user interface.  The feedback gains were normally small and the time constants were several to 100\,s.  The same procedure was also applied to stabilize hardware such as amplitudes and timings of power supplies.  While those loops were simple PI (Proportional-Integral) controllers, they were very effective.  

The fluctuations often meant that certain equipment or utility became out of order.  It was not determined well how much stability was required for such equipment at the beginning.  Because it may take time to analyze and fix those pieces of equipment, feedback loops were effectively applied at many locations.  

As the fluctuation sources were identified and resolved, those closed loops became unnecessary during the normal operation.  However, they are still important during the beam studies because the beam conditions are much different from the normal ones.  While a certain parameter is scanned in such studies, other parameters often have to be maintained stable.  They are also effective in order to identify the reason when the beam characteristics changed unintentionally.  Furthermore, they can be occasionally used to predict a certain failures.  Thus, the information from those loops was valuable. 

Those feedback loops were dependent on the beam modes.  The linac accelerates the beams for four storage rings, KEKB-HER, KEKB-LER, PF and PF-AR, with quite different energies.  Moreover, those beam modes had been exchanged every several minutes.  Even with such frequent beam mode changes the beam reproducibility and stabilities were maintained with the beam feedback loops.  


\section{Simultaneous Top-up Injections}

Until 2008 it took from 30\,seconds to 2\,minutes to switch the beam injection modes between KEKB-HER, KEKB-LER, PF and PF-AR.  Many accelerator device parameters were changed, and the slowest process was the bending magnet standardization.  However, simultaneous top-up injections to three rings, KEKB-HER, KEKB-LER and PF became necessary in order to improve the physics experiments at those rings. 

Several sets of pulsed equipment were installed.  Beam optics development was also performed to support the wide dynamic range of the beam energy and charge, 3-times different energies and 100-times different charges.  The betatron matching condition was established at the entrance to each beam transport line.  An event-based control system was introduced, in addition to existent EPICS control system, in order to achieve global and fast controls.  


The new control system could change many parameters quickly and globally, and it enabled the pulse-to-pulse beam modulation.  The linac pulse repetition is 50\,Hz, and approximately 130 independent parameters over 600-m linac are changed within 20\,ms.  Such efforts enabled simultaneous top-up injections, and the beam current stabilized to be 1\,mA in 1.1 $\sim$ 1.6\,A at KEKB and 0.1\,mA in 450\,mA at PF, which contributed the physics experiment results.  

\subsection{Event System Operation}

The new event-based control system arranges the linac for a pulse to operate in one of the ten beam modes.  The injection beam modes for KEKB-HER, KEKB-LER and PF can be switched quickly, and each beam pulse is separated by 20\,ms.  

The event-based control system is composed of three basic parts.  The first part is the beam-mode pattern arbitrator/generator.  It listens to the beam frequency requests from downstream rings and/or human operators, and then it arbitrates those requests based on pre-assigned priorities, and generates a beam mode pattern for up to 10\,s (20\,ms $\times$ 500).  There are certain constraints how to organize the pattern because many of the pulsed power supplies expect that they are fired at constant intervals. 

The main part is the event generator station.  The event generator (EVG230 from MRF~\cite{mrfweb}) provides an event sequence synchronized to RF clock (114.24\,MHz).  Each event is accompanied with an event code, while additional data is transferred as well.  The event generator software extracts adjacent beam-mode elements out of the beam-mode pattern received from the pattern generator, and arranges several event codes corresponding to the first mode, then adds another event code to notify the beam mode of the next pulse.  

\begin{figure}[tb]
   \centering
   \includegraphics*[angle=0, width=83mm]{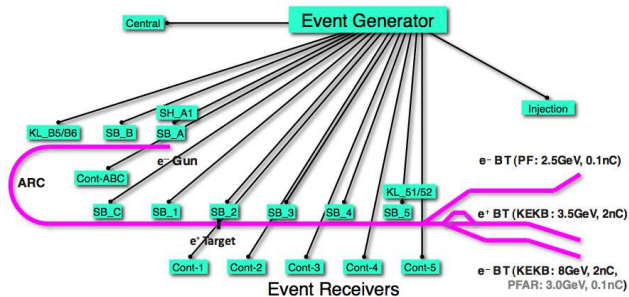}
   \caption{Overall configuration of the event-based control system.  17 event receiver stations cover 1-km facility.}
   \label{fig-config}
\end{figure}

Finally, event receiver stations accept events from the event generator through optical fiber links.  The event receiver (EVR230RF from MRF) regenerates the RF clock using the bit train.  The first part of the event code sequence is used to generate signals with specific delays.  The last event code informs the receiver software to prepare specific delay and analog values for the beam mode in the next pulse.  Approximately 130 parameters on 17 event receiver stations are changed every pulse (Fig.~\ref{fig-config}).  Each one of those parameters is associated with ten variables that correspond to ten beam modes, and those thousand variables can be manipulated any time by operational software. 

The beam-mode pattern arbitrator normally accepts the new requests from rings every several seconds, and it regenerates a corresponding pattern.  Under a typical operation condition, average injection rates are 25\,Hz for HER, 12.5\,Hz for LER and 0.5\,Hz for PF.  However, it is often required to assign all the pulses for injection.  Such a flexible injection pulse arrangement enables efficient use of the injector linac. 

\subsection{Beam Position Monitor}

As described previously, the old read-out system for beam position monitors (BPM) could operated at 1\,Hz.  However, for simultaneous injection it was required to process signals at 50\,Hz.  

The new BPM read-out system was designed with oscilloscopes, mainly because one oscilloscope can cover several BPMs and the system is simple with only passive components, so that the maintenance becomes easier.  The same software as the previous system was embedded into the EPICS control software framework on Windows XP on the oscilloscopes.  It accepts the event information through the network\footnote{Events are missed only less than once in million times}.  Events are used to tag the beam-position and -charge information with the beam-mode information.  As approximately 200 BPMs are installed along the linac and BT, each one of which has independent variables for ten beam modes, there are $\sim$2000 variables provided.   The client operational software can receive beam information which is related to a specific beam mode~\cite{linacbpm-ical09}.  

\section{Virtual Accelerators}

Under the simultaneous injection configuration the event-based control system provides beam-mode dependent control parameters.  Moreover, those parameters in different beam modes are organized to be independent both for controls and measurements.  Thus, we can see those independent parameter sets as independent virtual accelerators.  For each 20-ms time slot, the event system associates one of the virtual accelerators with the real accelerator. 

Because those control parameters for each virtual accelerator continue to exist, human operators and operational software can act on one of those virtual accelerators without any interference between other beam modes. 

\subsection{Event-based Beam Feedback Loops}

\begin{figure}[tb]
   \centering
   \includegraphics*[angle=0, width=83mm]{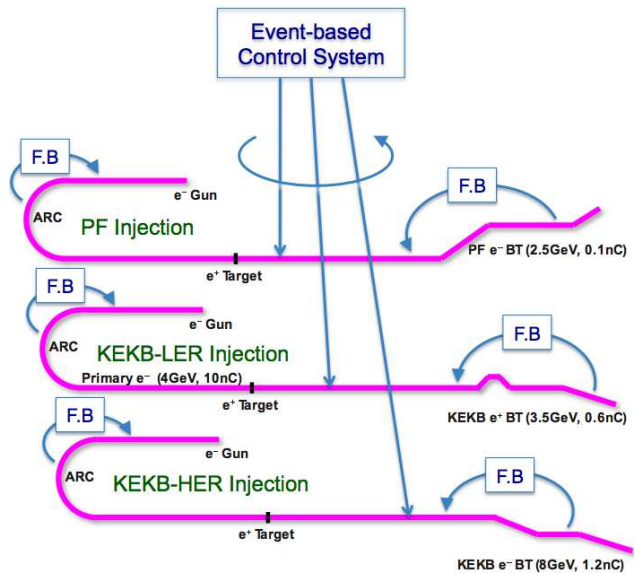}
   \caption{Independent closed feedback loops (F.B) on three virtual accelerators for KEKB-HER, KEKB-LER, PF.}
   \label{fig-virt}
\end{figure}

BPM information and RF control parameters are also handled independently in each virtual accelerator.  At first, energy feedback loops at the 180-degree arc and at the end of linac were installed again using event control parameters on each virtual accelerator as in Fig.~\ref{fig-virt}.  As parameters are independently managed, no modification to the software was necessary.  

The performance of those closed loops were observed with small feedback gains during the normal operation.  In those feedback operations no beam stability improvements were achieved.  In other words, no signs of instabilities were observed other than white noise.  

For PF BT energy stabilization, it turned out that the separation of betatron and dispersion functions was not optimal and the resolution of BPMs was insufficient because of the low beam charges.  The procedures of the betatron oscillation compensation and the weighed average of beam positions will be applied later.  Because the processing speed with a scripting language is not sufficient, compiled procedures are tested as well. 

The orbit and energy spread stabilizations can be implemented in the same way.  Those beam feedback signals will be valuable information for the accelerator operation.  

\section{Conclusion}

A pulse-to-pulse modulated simultaneous injection to KEKB-HER, KEKB-LER and PF rings was realized with an event-based fast and global control system.  It provides several virtual accelerators for the accelerator operations with independent parameter sets.  Under such environment, event-based closed beam energy feedback loops were successfully applied, which would provide valuable resources for the future operations including SuperKEKB.


\begin{thebibliography}{9}   

\bibitem{lin-status-linac02}
K.~Furukawa {\it et~al.},
``The Present Performance and Future Upgrade of the KEKB Electron Linac'',
{\em Proc.\ LINAC2002}, Gyeongju, Korea, 2002, TU431, p.~382.

\bibitem{beamswitch-linac00}
K.~Furukawa {\it et~al.},
``Beam Switching and Beam Feedback Systems at KEKB Linac'',
{\em Proc.\ LINAC2000}, Monterey, USA, 2000, TUE10, p.~633.

\bibitem{lin-efb-ical99}
K.~Furukawa {\it et~al.},
``Energy Feedback Systems at KEKB Injector Linac'',
{\em Proc.\ ICALEPCS'99}, Trieste, Italy, 1999, TA1O04, p.~248.


\bibitem{iuc-linac08}
M.~Satoh {\it et~al.},
``Present Status of the KEK Injector Upgrade for the Fast Beam-mode Switch'',
{\em Proc.\ LINAC2008}, Vancouver, Canada, 2008, TUP013, p.~416.

\bibitem{eventcont-ical09}
K.~Furukawa {\it et~al.},
``New Event-based Control System for Simultaneous Top-up Operation at KEKB and PF'',
{\em Proc.\ ICALEPCS2009}, Kobe, Japan, 2009, THP052. 

\bibitem{simul-pac09}
N.~Iida {\it et~al.}, 
``Pulse-to-Pulse Switching Injections to Three Rings of Different Energies from a Single Electron Linac at KEK'', 
{\em Proc.\ PAC2009}, Vancouver, Canada, 2009, WE6PFP110.

\bibitem{mrfweb}
\texttt{<http://www.mrf.fi/>}.

\bibitem{linacbpm-ical09}
M.~Satoh {\it et~al.},
``EPICS IOC of WindowsXP-based Oscilloscope for Fast BPM Data Acquisition System'',
{\em Proc.\ ICALEPCS2009}, Kobe, Japan, 2009, WEP086. 




\end{thebibliography}
\end{document}